\documentclass[conference]{IEEEtran}
\IEEEoverridecommandlockouts
\usepackage{cite}
\usepackage{amsmath,amssymb,amsfonts}
\usepackage{algorithmic}
\usepackage{graphicx}
\usepackage{textcomp}
\usepackage{xcolor}
\def\BibTeX{{\rm B\kern-.05em{\sc i\kern-.025em b}\kern-.08em
    T\kern-.1667em\lower.7ex\hbox{E}\kern-.125emX}}
\begin{document}

\newcommand{\vlad}[1]   {\noindent \textcolor{cyan}{[#1 -Vlad]}}
\newcommand{\tommi}[1]   {\noindent \textcolor{magenta}{[#1 -Tommi]}}

\title{Software Architecture Challenges in Integrating Hybrid Classical-Quantum Systems}

\author{
\IEEEauthorblockN{Vlad Stirbu}
\IEEEauthorblockA{
\textit{University of Jyväskylä}\\
Jyväskylä, Finland \\
vlad.a.stirbu@jyu.fi}
\and
\IEEEauthorblockN{Tommi Mikkonen}
\IEEEauthorblockA{
\textit{University of Jyväskylä}\\
Jyväskylä, Finland \\
tommi.j.mikkonen@jyu.fi}
}

\maketitle

\begin{abstract}
The emergence of quantum computing proposes a revolutionary paradigm that can radically transform numerous scientific and industrial application domains. The ability of quantum computers to scale computations exponentially imply better performance and efficiency for certain algorithmic tasks than current computers provide. However, to gain benefit from such improvement, quantum computers must be integrated with existing software systems, a process that is not straightforward. In this paper, we investigate challenges that emerge from building larger hybrid classical-quantum computers, and discuss some approaches that could be employed to overcome these challenges.
\end{abstract}


\section{Introduction}


Quantum computers have demonstrated the potential to revolutionize various fields, including cryptography, drug discovery, materials science, and machine learning, by leveraging the principles of quantum mechanics. However, the current generation of quantum computers, known as noisy intermediate-scale quantum (NISQ) computers, suffer from noise and errors, making them challenging to operate. Additionally, the development of quantum algorithms requires specialized knowledge not readily available to the majority of software professionals. These factors pose a significant entry barrier for leveraging the unique capabilities of quantum systems.

For the existing base of business applications, classical computing has already proven its capabilities across a diverse range of solutions. However, some of the computations they must perform can be accelerated with quantum computing, much like GPUs are used today. Therefore, quantum systems should not function in isolation, but they must coexist and interoperate with classical systems. To this end, software architects play a crucial role in achieving seamless integration, while simultaneously designing systems that effectively meet the unique requirements of businesses.

To address the challenges associated with this integration, this paper focuses on designing hybrid systems that integrate quantum and classical computing, aiming to overcome architectural, design, and operational hurdles. In doing so, we look at the software development lifestyle, the technology stack of hybrid classic-quantum systems, and deployment techniques used today. 

\section{Background}

\begin{figure*}[t]
    \centering
    \includegraphics[width=0.8\textwidth]{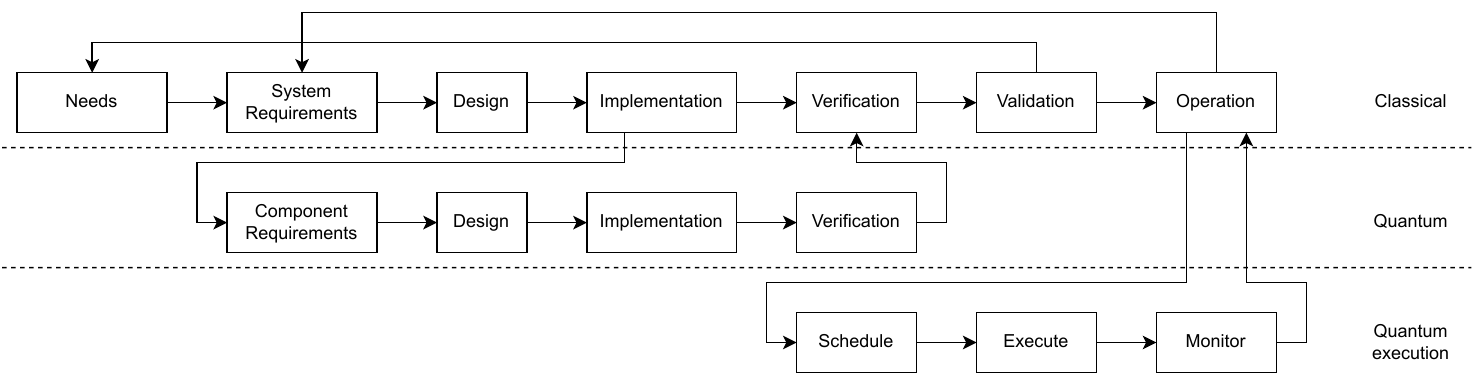}
    \caption{Software development lifecycle of a hybrid classical-quantum system}
    \label{fig:sdlc}
\end{figure*}

The software development lifecycle (SDLC) of hybrid classic-quantum applications consist of a multi-faceted approach, as depicted in Fig. \ref{fig:sdlc}. At the top level, the classical software development process starts by identifying user needs and deriving them into system requirements. These requirements are transformed into a design and implemented. The result is verified against the requirements and validated against user needs. Once the software system enters the operational phase, any detected anomalies are used to identify potential new system requirements, if necessary. A dedicated track for quantum components is followed within the SDLC \cite{sdlc}, specific to the implementation of quantum technology. The requirements for these components are converted into a design, which is subsequently implemented on classic computers, verified on simulators or real quantum hardware, and integrated into the larger software system. During the operational phase, the quantum software components are executed on real hardware. Scheduling ensures efficient utilization of scarce quantum hardware, while monitoring capabilities enable the detection of anomalies throughout the process.

\begin{figure}
    \centering
    \includegraphics[width=0.4\textwidth]{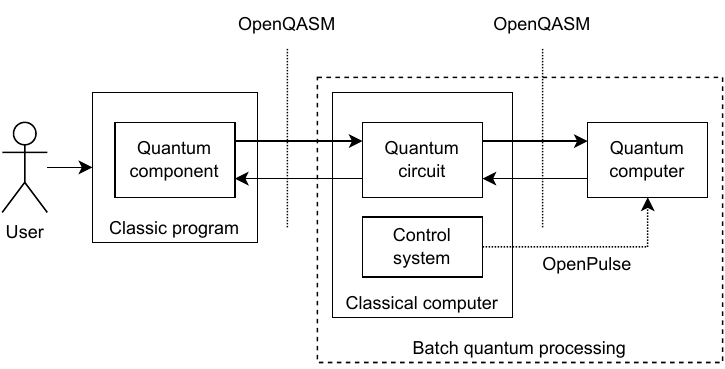}
    \caption{Quantum computing model: components and interfaces}
    \label{fig:system}
\end{figure}

A typical hybrid classic-quantum software system is understood as a classical program that has one or more software components that are implemented using quantum technology, as depicted in Fig. \ref{fig:system}. A quantum component utilises quantum algorithms \cite{Montanaro2016}, that are transformed into quantum circuits using a toolkit like Cirq\footnote{https://quantumai.google/cirq} or Qiskit\footnote{https://qiskit.org}. The quantum circuit describes quantum computations in a machine-independent language using quantum assembly (QASM) \cite{openqasm}. This circuit is translated by a computer that controls the quantum computer in a machine specific circuit and a sequence of pulses that control the operation of individual hardware qubits \cite{openpulse}. Due to the scarcity or quantum hardware and the process of preparing the individual runs, the quantum task execution process is lengthy, having the characteristics of batch processing in classical computing. In fact, techniques used in batch processing, such as Slurm \cite{yoo2003slurm}, can be used to implement this step, which adds indirection to the underlying software architecture.

\section{Architectural concerns}


\subsection{Design -- Algorithms, data structures and APIs}

Quantum algorithms are designed specifically to take advantage of quantum mechanics properties such as quantum superposition and entanglement. They provide advantages over classical equivalents for specific areas, such as factoring or linear search. Software architects should evaluate the feasibility to achieve quantum advantage during the component requirements phase of the SDLC. They must ensure that the needed computational resources are available and that data can be mapped from the classic to quantum domains. For example, TensorFlow Quantum\footnote{https://www.tensorflow.org/quantum} is a library for rapid prototyping of hybrid quantum-classical ML models that focuses on quantum data and hybrid quantum-classical machine learning models.

The batch nature of the quantum task execution has a profound impact on the software architecture of a hybrid classic-quantum system. The jobs submitted for execution are queued and scheduled using fair-share policies. As the task execution results are not available immediately, the software system should favour  \textit{asynchronous} communication. Further, the system designers must consider the \textit{security and privacy} aspects of executing tasks on quantum hardware infrastructure shared by several organizations.



\subsection{Operations -- Implementation leak and resource allocation}

Quantum application written using popular libraries like Cisq and Qiskit have a \textit{monolith} nature. They combine into a single \textit{imperative} program the application logic components, the general purpose quantum circuit design, the quantum hardware selection (e.g. backend configuration), and the transformation of the machine specific circuit that is actually executed. To make the software architecture modular, the general purpose part needs to be separated from the quantum backend. Essential backend information, such as the quantum volume (the qubit connectedness), needs to be accessed at runtime so that the actual hardware selection can be done dynamically based on dynamic factors, like hardware availability (if there are multiple providers) and cost estimates. For example, Kubernetes serves as an extensible orchestration platform that enables efficient scheduling of classic computing jobs. The capabilities of quantum computers can be exposed in this computing environment, while the scheduler can be enhanced to efficiently handle quantum jobs.



\section{Conclusions and future steps}
The fundamental differences in programming models and the varying levels of maturity in tools and practices between the classical and quantum domains makes their seamless integration difficult. To gain insights and firsthand experience, we intend to collaborate with the users of HELMI\footnote{https://docs.csc.fi/computing/quantum-computing/overview/} quantum computer, in an effort to overcome the integration barriers between classical and quantum computing.


\section*{Acknowledgement}
This work has been supported by the Academy of Finland (project DEQSE 349945) and Business Finland (project TORQS 8582/31/2022).

\bibliographystyle{ieeetr}
\bibliography{references}

\begin{thebibliography}{1}

\bibitem{sdlc}
B.~Weder, J.~Barzen, F.~Leymann, and D.~Vietz, {\em Quantum Software
  Development Lifecycle}, pp.~61--83.
\newblock Cham: Springer International Publishing, 2022.

\bibitem{Montanaro2016}
A.~Montanaro, ``Quantum algorithms: an overview,'' {\em npj Quantum
  Information}, vol.~2, p.~15023, Jan 2016.

\bibitem{openqasm}
A.~Cross, A.~Javadi-Abhari, T.~Alexander, N.~De~Beaudrap, L.~S. Bishop,
  S.~Heidel, C.~A. Ryan, P.~Sivarajah, J.~Smolin, J.~M. Gambetta, and B.~R.
  Johnson, ``Openqasm 3: A broader and deeper quantum assembly language,'' {\em
  ACM Transactions on Quantum Computing}, vol.~3, sep 2022.

\bibitem{openpulse}
T.~Alexander, N.~Kanazawa, D.~J. Egger, L.~Capelluto, C.~J. Wood,
  A.~Javadi-Abhari, and D.~C. McKay, ``Qiskit pulse: programming quantum
  computers through the cloud with pulses,'' {\em Quantum Science and
  Technology}, vol.~5, p.~044006, aug 2020.

\bibitem{yoo2003slurm}
A.~B. Yoo, M.~A. Jette, and M.~Grondona, ``{Slurm: Simple Linux utility for
  resource management},'' in {\em Job Scheduling Strategies for Parallel
  Processing: 9th International Workshop, JSSPP 2003, Seattle, WA, USA, June
  24, 2003. Revised Paper 9}, pp.~44--60, Springer, 2003.

\end{thebibliography}
\end{document}